\documentstyle{mn}
\input{psfig}
\newif\ifAMStwofonts

\def\simgt{\hbox{\rlap{\raise 0.425ex\hbox{$>$}}\lower 0.65ex\hbox{$\sim$}}}
\def\simlt{\hbox{\rlap{\raise 0.425ex\hbox{$<$}}\lower 0.65ex\hbox{$\sim$}}}

\ifoldfss
  \ifCUPmtlplainloaded \else
    \NewTextAlphabet{textbfit} {cmbxti10} {}
    \NewTextAlphabet{textbfss} {cmssbx10} {}
    \NewMathAlphabet{mathbfit} {cmbxti10} {} 
    \NewMathAlphabet{mathbfss} {cmssbx10} {} 
  \fi
  \ifAMStwofonts
    \ifCUPmtlplainloaded \else
      \NewSymbolFont{upmath} {eurm10}
      \NewSymbolFont{AMSa} {msam10}
      \NewMathSymbol{\upi}     {0}{upmath}{19}
      \NewMathSymbol{\umu}     {0}{upmath}{16}
      \NewMathSymbol{\upartial}{0}{upmath}{40}
      \NewMathSymbol{\leqslant}{3}{AMSa}{36}
      \NewMathSymbol{\geqslant}{3}{AMSa}{3E}

      \let\leq=\leqslant 
       
    \fi
  \fi
\fi 

\ifnfssone
  \newmathalphabet{\mathit}
  \addtoversion{normal}{\mathit}{cmr}{m}{it}
  \addtoversion{bold}{\mathit}{cmr}{bx}{it}
  \newmathalphabet{\mathbfit} 
  \addtoversion{normal}{\mathbfit}{cmr}{bx}{it}
  \addtoversion{bold}{\mathbfit}{cmr}{bx}{it}
  \newmathalphabet{\mathbfss} 
  \addtoversion{normal}{\mathbfss}{cmss}{bx}{n}
  \addtoversion{bold}{\mathbfss}{cmss}{bx}{n}
  \ifAMStwofonts
    \ifCUPmtlplainloaded \else
      \UseAMStwoboldmath
      \makeatletter
      \new@mathgroup\upmath@group
      \define@mathgroup\mv@normal\upmath@group{eur}{m}{n}
      \define@mathgroup\mv@bold\upmath@group{eur}{b}{n}
      \edef\UPM{\hexnumber\upmath@group}
      \new@mathgroup\amsa@group
      \define@mathgroup\mv@normal\amsa@group{msa}{m}{n}
      \define@mathgroup\mv@bold\amsa@group{msa}{m}{n}
      \edef\AMSa{\hexnumber\amsa@group}
      \makeatother
      \mathchardef\upi="0\UPM19
      \mathchardef\umu="0\UPM16
      \mathchardef\upartial="0\UPM40
      \mathchardef\leqslant="3\AMSa36
      \mathchardef\geqslant="3\AMSa3E

      \let\leq=\leqslant 

    \fi
  \fi
\fi 

\ifnfsstwo
  \DeclareMathAlphabet{\mathbfit}{OT1}{cmr}{bx}{it}
  \SetMathAlphabet\mathbfit{bold}{OT1}{cmr}{bx}{it}
  \DeclareMathAlphabet{\mathbfss}{OT1}{cmss}{bx}{n}
  \SetMathAlphabet\mathbfss{bold}{OT1}{cmss}{bx}{n}
  \ifAMStwofonts
    \ifCUPmtlplainloaded \else
      \DeclareSymbolFont{UPM}{U}{eur}{m}{n}
      \SetSymbolFont{UPM}{bold}{U}{eur}{b}{n}
      \DeclareSymbolFont{AMSa}{U}{msa}{m}{n}
      \DeclareMathSymbol{\upi}{0}{UPM}{"19}
      \DeclareMathSymbol{\umu}{0}{UPM}{"16}
      \DeclareMathSymbol{\upartial}{0}{UPM}{"40}
      \DeclareMathSymbol{\leqslant}{3}{AMSa}{"36}
      \DeclareMathSymbol{\geqslant}{3}{AMSa}{"3E}

      \let\leq=\leqslant 

    \fi
  \fi
\fi 

\ifCUPmtlplainloaded \else
  \ifAMStwofonts \else 
    \def\upi{\pi}
    \def\umu{\mu}
    \def\upartial{\partial}
  \fi
\fi

\def \Om {\Omega_0}
\def \lo {\lambda_0}
\def \bj {b_{{\rm J}}}
\def \mb {M_{\rm b}}
\def \kms {{\rm km\,s}$^{-1}$}
\def \lam {$\lambda$}

\def \lb {L_{\rm B}}
\def \mb {M_{\rm B}}
\def \mbh {M_{\rm BH}}
\def \lya {Ly$\alpha$}
\def \lyb {Ly$\beta$}
\def \ha {H$\alpha$}
\def \hb {H$\beta$}
\def \hg {H$\gamma$}
\def \hd {H$\delta$}
\def \oiii {[O{\small~III}]}
\def \oii {[O{\small~II}]}

\def \neiii {[Ne{\small~III}]}
\def \nev {[Ne{\small~V}]}
\def \mgii {Mg{\small~II}}

\def \civ {C{\small~IV}}
\def \ciii {C{\small~III}]}
\def \siiv {Si{\small~IV}}
\def \oiv {O{\small~IV}}
\def \feii {Fe{\small~II}}
\def \nv {N{\small~V}}

\def \aliii {Al{\small~III}}
\def \siiv {Si{\small~IV}}
\def \siiii {Si{\small~III}]}

\def \rhb {r_{\rm H\beta}}

\def \mnras {MNRAS}
\def \apj {ApJ}

\def \aap {A\&A}

\title[Line wdiths and QSO black hole masses]{Emission line widths and
QSO black hole mass estimates from the 2dF QSO Redshift Survey.} 
\author[E. A. Corbett et al.]
{E. A. Corbett$^1$\thanks{ecorbett@aaoepp.aao.gov.au},
  S. M. Croom$^1$, B. J. Boyle$^1$, H. Netzer$^2$,
\newauthor L. Miller$^3$, P. J. Outram$^4$, T. Shanks$^4$, R. J. Smith$^5$, K. Rhook$^{1,6}$ \\
$^1$ Anglo-Australian Observatory, PO Box 296, Epping, NSW 1710, Australia\\
$^2$ School of Physics and Astronomy, Tel-Aviv University, Tel-Aviv
69978, Israel\\
$^3$Department of Physics, Oxford University, Keble Road, Oxford, OX1
3RH, UK\\
$^4$Physics Department, University of Durham, South Road, Durham, DH1 3LE,
UK\\
$^5$Astrophysics Research Institute, Liverpool John Moores University,
Twelve Quays House, Egerton Wharf, Birkenhead,  CH41 1lD, UK\\
$^6$ School of Physics, University of Melbourne, VIC 3010, Australia
}
\begin{document}

\maketitle

\begin{abstract}

We have used composite spectra generated from more than 22000 QSOs
observed in the course of the 2dF and 6dF QSO Redshift Surveys to
investigate the relationship between the velocity width of emission
lines and QSO luminosity.  We find that the velocity width of the
broad emission lines \hb, \hg, \mgii, \ciii\ and \civ\ are correlated
with the continuum luminosity, with a significance of more than 99 per
cent. Of the major narrow emission lines (\oiii\ \lam5007, \oii\
\lam3727, \neiii\ \lam3870 and \nev\ \lam3426) only \oiii\ exhibits a
significant correlation between line width and luminosity. Assuming
that the gas is moving in Keplerian orbits and that the radius of the
broad line region is related to the QSO continuum luminosity, we use
the velocity widths of the broad lines to derive average black hole
masses for the QSOs contributing to the composite spectra. The
resultant QSO mass-luminosity relationship is consistent with
$M\propto L^{0.97\pm0.16}$. We find that the correlation between line
width and redshift, if present, must be weak, and only \civ\ shows
significant evidence of evolution. This enables us to constrain the
redshift evolution of the black hole mass-luminosity ratio to be
$\sim(1+z)^\beta$ with $\beta\simlt1$, much less than the
$\sim(1+z)^3$ evolution seen in QSO luminosity evolution. Assuming
that the motion of the broad line region gas is Keplerian and that its
radius depends on the QSO luminosity, our models indicate that the
observed weak redshift dependence is too small for the observed QSO
luminosity function to be due to the evolution of a single long-lived
population of sources.

 \end{abstract}


\section{Introduction}

The vast energy budget required by active galactic nuclei (AGN) is
probably provided by the accretion of matter onto a supermassive black
hole (BH). Evidence that such objects can be found in AGN is provided by
the Keplerian motions of the mega-masers observed in some sources
(e.g. Miyoshi et al. 1995),  \nocite{Miyoshi95} the broad FeK$\alpha$
emission line observed in the X-ray spectrum of some sources (e.g.
MCG-6-30-15; see Tanaka et al. 1995; Wilms et al. 2001)
\nocite{Tanaka95} \nocite{Wilms01} and the reverberation mapping
experiments (e.g. Blanford \& McKee 1982; Netzer  \& Peterson 1997;
Wandel, Peterson \& Malkan 1999; Kaspi et al. 2000, hereafter K00;
Peterson et al. 2000).  \nocite{Blanford82}\nocite{Peterson00}
\nocite{Wandel99}\nocite{Kaspi96}\nocite{Kaspi00} The gas in the broad
line region (BLR) resides close to the central continuum source and
can therefore be used as a probe of the central mass. In particular,
if the BLR is gravitationally bound and in near-Keplerian orbits, it
should be possible to estimate the  central mass, $M$, from the mean
radius, $r$, and the velocity dispersion, $v$, of the emitting gas.

Evidence that the velocity dispersions of several emission lines, in
particular \hb, are indeed dominated by Keplerian motions is presented
by Peterson \& Wandel (1999) and Onken \& Peterson (2002). Velocity
widths measured either from the rms spectra or the mean spectrum,
combined with reverberation mapping measurements of the BLR radius,
are now available for more than 34 sources  (K00, and references
therein; Onken \& Peterson 2002; Woo \& Urry 2002, and references
therein)\nocite{Kaspi00}.  Although there are some problems with this
method (see Krolick 2001 for a detailed discussion), comparison of
masses obtained in this way  with those obtained from the bulge
stellar velocity dispersion have been found to be in good agreement
(Ferrarese et al. 2001\nocite{Ferrarese01}, Gebhardt et
al. 2000\nocite{Gebhardt00}, Woo \& Urry 2002\nocite{Woo02}).

Reverberation mapping is extremely time consuming, requiring that
source light curves are well sampled over a number of years and that
the sources themselves are variable. As a result of these constraints,
reverberation mapping has only been used to estimate the virial mass
in a small number ($<$40) of relatively low redshift objects.  It is
not practical to expand these studies to obtain masses for a larger
number of sources spanning a larger range in redshift and luminosity.

The available reverberation mapping results  show a clear correlation
 between the radius of the \hb\ emitting region, $\rhb$, and the
 monochromatic luminosity at 5100\AA\ such that $\rhb\propto \lambda
 L^{\gamma}_{5100}$
 \cite{Kaspi00,Peterson00,Mclure02,Vestergaard02,Netzer03} where $0.5
 < \gamma < 0.7$. The uncertainty on the slope $\gamma$ reflects the
 large scatter in the data as well as the statistical method used for
 the linear regression. These issues have been discussed, in detail,
 by Vestergaard (2002), McLure and Jarvis (2002), Maoz (2002) and
 others. The most recent paper on this topic, by Netzer (2003), also
 addresses these issues and comments on the use of the so called
 ``photoionization radius'' (see Wandel et
 al. 1999). \nocite{Wandel99} Regardless of the exact value of
 $\gamma$, such a relationship can be used to estimate $\rhb$ in {\it
 every source} of known luminosity and to obtain the BH mass by
 combining this size with the observed \hb\ line width. This method
 has been referred to as the {\it single epoch mass estimate}.

Reverberation mapping experiments have concentrated on the \hb\
emission line.  This has restricted the use of the ``photionization
method'' to the measurement of BH mass in relatively low redshifts
($z<$0.9) and low luminosity objects, although there are some cases of
high-z quasars where the \hb\ line has been observed in the infrared
(Shields et al. 2002 and references therein).  The obvious extension
to high redshift sources is to use the UV luminosity as an estimate of
$r$ and the width of a certain  UV line (e.g. \mgii, \civ\ or  \lya)
as a measure of the gas velocity. Unlike the case of \hb, where the
location of the line emission ($\rhb$) is known, there is no
direct measure of the location of the UV lines except in a handful of
sources. Thus, the only solution is to calibrate the known
relationship for \hb\ against other lines assuming the two must result
in the same derived BH mass. This must be done for sources where
$\rhb$ is known and where the velocity width of the UV line and
the UV continuum luminosity can be measured.  Such intercalibration
has been performed, successfully, for \civ\ by Vestergaard
(2002)\nocite{Vestergaard02} and for \mgii\ by McLure \& Jarvis
(2002)\nocite{Mclure02}. Thus, both \civ\ and \mgii\ can be used to
obtain single-epoch BH  masses.

Although the results from both Vestergaard (2002) and Mclure \& Jarvis
(2002) are encouraging, their correlations display a large amount of
scatter. This may well be intrinsic, rather than due to poor signal to
noise. One possibility (Young et al. 1999; McLure \& Dunlop 2002;
Smith et al. 2002)\nocite{Mclure02a}\nocite{Smith02}\nocite{Young99}
is that some (or even most) BLRs  display a flattened disc-like
geometry.  In this case, the emission line widths would  be a function
of the viewing angle, with larger widths observed at larger
inclinations, resulting in the velocity dispersion (and hence mass) of
a face-on system being underestimated. In addition, the K00 original
relationship which is the basis of all further work, is limited to
$\lambda L_{\lambda} (5100\AA) < 10^{46}$ ergs~s$^{-1}$. Thus, mass
estimates for the highest luminosity sources are uncertain because of
the necessary extrapolation.

In this paper, we make use of high signal-to-noise ratio ($SNR$)
composite spectra from the 2dF QSO Redshift Survey 2QZ (2QZ; see Croom
et al.\ 2002) to test the  velocity-luminosity relationship in QSOs
and to obtain new BH mass estimates. The use of composite spectra
minimizes the scatter in the relation between luminosity and velocity
width caused by effects such as inclination angle. It also minimizes
the influence of objects with extremely small or extremely large line
widths. The broad range in luminosity and redshift allows us to
investigate correlations of velocity width and BH mass with  redshift
and luminosity.

In Section 2 we briefly describe the primary spectral database and
outline the procedure used to obtain the composite spectra. We discuss
the  method used to measure the emission line widths in Section 3 and
derive relations between velocity width and luminosity for a number
of emission lines in Section 4. In Section 5 we extend the work to
estimate black hole masses from the UV broad emission line widths and
the QSO luminosity, calibrating them against the reverberation mapping
results for \hb,  and discuss the implications of these results for
both QSO evolution and  BLR structure. Our conclusions are summarized
in Section 6.

\begin{figure*}
\centering \centerline{\psfig{file=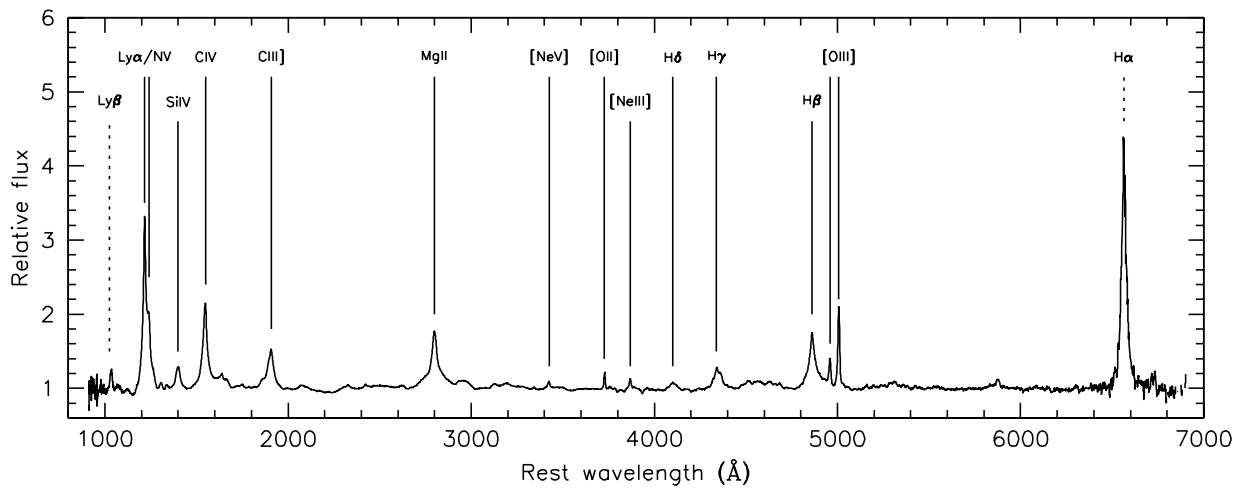,width=18cm}}
\caption{The QSO composite spectrum composed of all the QSOs used in
our analysis. Emission lines labeled with solid lines were included in
our study, lines labeled with dotted lines were excluded.}
\label{fig_compall}
\end{figure*}

\section{Data}

\subsection{Observations}

The data used in our analysis is taken from the 2dF and 6dF QSO
Redshift Surveys (2QZ, Croom et al.\ 2001; 6QZ Croom et al. 2003, in
preparation).  QSO candidates were selected from UK Schmidt
photographic material over the range $16.0<\bj\leq20.85$ using their
$u\bj r$ colours.  Photometric errors in the $\bj$ band were typically
0.1 mag.  Objects brighter than $\bj=18.25$ were observed with the 6dF
facility on the UK Schmidt over the range 3900--7600\AA\ with a
spectral resolution of 11\AA. Those fainter than $\bj=18.25$ were
observed with 2dF on the AAT with a spectral range of 3700--7900\AA\
and a spectral resolution of 9\AA.  Typical redshift errors are
$\sigma_{\rm z}=0.003$. The median SNR was 15 for the 6dF spectra and
5 for the 2dF spectra. In total 22041 independent QSO spectra from the
2dF and 6dF observations were included in the dataset.

Absolute magnitudes ($\mb$) were computed from the observed
photographic $\bj$ magnitude, after correction for Galactic extinction
\cite{sfd98}, using the K-corrections found by Cristiani \& Vio
(1990)\nocite{cv90}.  Throughout we assume a flat cosmological world
model with $\Om=0.3$, $\lo=0.7$ and $H_0=70$~\kms~Mpc$^{-1}$.

\subsection{Composite spectra}

Full details of the creation of the composite spectra and line
profile fitting is given by Croom et al. (2002). Briefly, composite
QSO spectra (excluding BAL QSOs) were created in discrete absolute
magnitude ($\Delta\mb=0.5\,$mag) and redshift ($\Delta z = 0.25$)
bins. A further series of composite spectra were obtained from the QSOs
binned in absolute magnitude only, as well as a composite contained all
QSOs.  The numbers of QSOs contributing to each composite are given in
Table 1 in Croom et al. (2002), typically ranging from 100 to 1000
objects.

The individual spectra were not flux calibrated and each individual
spectrum was normalised by a continuum fitted between known emission
features.  Due to the limited wavelength coverage of these spectra it
was necessary to treat the large \feii\ features underlying the
\mgii\ \lam2798 emission line as continuum. After continuum
normalization the spectra were shifted to the rest frame,
interpolating linearly onto a uniform scale of 1\AA\,bin$^{-1}$.
Finally the composite spectrum in each $\mb,z$ bin was produced by
taking the median value in each wavelength element. The median $z$ and $\mb$ of the contributing QSOs was also determined for each wavelength element

The composite formed from all 22041 QSOs is shown in Figure 1. A
number of emission features can be clearly identified, including 3
narrow (forbidden) lines, 9 broad (permitted) emission lines and one
semi-forbidden line (\ciii\ \lam1909). The features identified in
Figure 1 with solid lines are suitable for detailed study because they
exhibit large equivalent widths (e.g. \lya, \civ\ \lam1909), are
relatively free from contamination by other emission lines and are
visible over a large redshift range. \ha\ or \lyb\ (Fig.1; dotted
lines) were not included in our study as, due to the redshift
distribution of the sample, too few spectra contain these emission
lines for meaningful conclusions to be drawn.

\subsection{Limitations}

Although the use of composite spectra provides a good picture of the
ensemble average properities of the QSO population, there may be small
subsamples of the population e.g. weak-lined QSOs for which the any
general trends identified below do not apply.

The use of composite spectra may also hide magnitude-dependent biases
against QSOs with, in particular, weak broad emission lines which
would prove more difficult to identify in the original low SNR
spectra.  Fortunately, although the spectroscopic identification of
QSOs in the 2QZ is a function of apparent magnitude, the
identification rate is generally high.  Croom et al. (2003) provide
an empirical fit to the overall spectroscopic identification rate
($f_{\rm s}$) as a function of apparent $\bj$ magnitude:
\begin{equation}
f_{\rm s}=1-\frac{\exp(\bj-20.39)}{(1-0.89)^{-0.92}}
\end{equation}
For the 6QZ survey, the spectroscopic identification rate is
more-or-less constant at $f_{\rm s}=0.97$.  Thus the identification
rate for the survey remains above 90 per cent for all but the faintest
$0.75\,$mag range of the combined sample.
\footnote {Based on repeat observations of sources for which we
initially failed to secure a spectroscopic identification, we also
confirmed that the fraction of QSOs amongst the unidentified sources
is similar to the fraction of QSOs amongst the identified sources.}
Even for the faintest magnitude, the identification rate does not go
below 80 per cent.

Finally, we note that as redshift increases, the observed spectrum
corresponds to a different portion of the rest-frame spectrum.  The
continuum normalization of the individual spectra that make up the
composite will therefore be defined by different regions of
rest-wavelength.  This has been largely circumvented by fixing
continuum bands in rest-wavelength near the lines of interest and
interpolating between them to define the continuum beneath emission
lines. Comparison of composite spectra generated for different
redshift bins within the same luminosity range do not reveal any
obvious evidence of such systematic redshift effects.  It is possible,
however, that low level systematic redshift biases are present in our
data, but we do not believe that they would be sufficiently strong to
alter our main conclusions.

\section{Emission line widths}

\subsection{Fitting}

The majority of the strong emission lines in the composite spectra
used in this study are blended with other weaker emission lines,
usually from different elements. Additionally, permitted emission
lines such as \hb\ and \hg\ often exhibit both a broad and a narrow
component which are emitted from physically distinct regions and we
therefore attempted to separate these components.

To measure line widths, the local ``pseudo-continuum'' on either side
of each feature was fitted with a straight line and subtracted. We
modeled the overlapping lines contributing to each spectral feature
using multi-component Gaussian fits. We note, however, that assuming a
Gaussian form for the features in our spectra is a gross
over-simplification, and thus the Gaussian fits were only used to
remove contaminating emission (which represented only a fraction of
the flux in the main line) with non-parametric methods used to
determine the line equivalent width, central wavelength and velocity
widths. For the purposes of this fitting process the narrow emission
lines and the broad Balmer lines (\hd, \hg\ and \hb) were modeled as
single Gaussians. When possible the number of independent parameters
in the model was reduced by linking some together. For example, since
the \oiii\ \lam\lam5007,4959 and narrow \hb\ emission arises from the
same region of the QSO (the narrow line region, NLR) it seems reasonable
to assume that the emitting gas will have similar velocity shifts and
dispersions.

The broad ultra-violet lines, i.e. from \mgii\ \lam2798 blue-ward,
display emission line profiles with very broad bases which cannot be
adequately modeled by a single Gaussian (Figure 1). They were
therefore fitted with two components; a very broad Gaussian (full
width at half maximum, FWHM$\sim10000$ \kms) and a narrower component
(FWHM $\sim2000-4000$ \kms). These components were constrained to
have the same central wavelength, with the exception of \lya\ as
absorption to the blue side of the line results in an asymmetric
profile and it was therefore necessary to allow a velocity shift
between the two components to fit the line profile. In all cases the
best fit to the spectral feature was found using $\chi^{2}$
minimization techniques. See Croom et al. (2002) for further details
on the line fitting procedure.

No attempt was made to model and remove the broad \feii\ emission
underlying the \mgii\ feature. Our normalisation procedure treats much
of this feature as continuum emission, since there is no true
continuum region close to the \mgii\ line, and it is therefore not
possible to model and remove it from the normalised spectrum. The
small \feii\ feature at $\sim2900$\AA\ just to the red of \mgii\ line,
was modeled and removed. It was also not possible to deblend the
\feii\ emission lines underlying the \hb\ spectral feature, the \oiv]
\lam1402 emission lines from the \siiv\ \lam1398 emission line and the
\siiii]\ \lam1892 emission from the \ciii\ \lam1909 feature (although
the \aliii\ \lam1857 emission was modeled and removed from \ciii).

Finally, the broad Balmer emission line \hg\ proved difficult to
deblend as it is contaminated by emission from both [\feii] \lam4358
and \oiii\ \lam4364 as well as narrow \hg\ emission. Since the
[\feii] and \oiii\ emission are within 6\AA\ of each other they are
not resolved in the 2dF spectra and were therefore modeled as single
narrow component centered between the two lines. The fit was further
constrained by fixing the velocity width of the narrow \hg\ and the
combined [\feii] and \oiii\ lines to that obtained for the \oiii\
\lam5007 emission.

Once we had modeled a spectral feature, the fits to the contaminating
line emission were subtracted, leaving only the line of
interest. Non-parametric methods (rather than the Gaussian fit) were
used to measure the flux, velocity widths and central wavelengths of
the emission line. These measurements were obtained for each feature
both before and after the contaminating emission was subtracted in
order to check for consistency.

\subsection{Measurement of line widths}

The emission lines widths were measured for the composites binned by
magnitude and redshift (76 spectra in total; $M-z$ composites
hereinafter). Line widths have been measured using several different
methods. These methods can be divided into two broad categories. The
most common method (e.g. Boroson \& Green 1992) is based on the full
width of the line at some fraction of its maximum flux, e.g. the full
width at half maximum (FWHM) or full width at zero intensity (FWZI),
whereas the second method measures the velocity width outside which
some fraction of the line flux falls, e.g.inter-percentile velocity
widths (IPVs, e.g. Whittle 1985; Stirpe 1991; Stirpe, Robinson \& Axon
1999)\nocite{Whittle85}\nocite{Stirpe91}. The most obvious difference
between these two methods is that in the former method the emphasis is
placed on the velocity width of the gas contributing to the core of
the line emission whereas the latter method takes into account the
total flux in the line. The relative merits of the two methods are
discussed elsewhere (see, e.g. Robinson 1995; Stirpe 1991) and it is not clear which gives the most consistent estimate of the emitting gas velocity. For this study we measured the emission
line width of our sample using both methods and comparison of the
velocity widths measured for the same line using the two sytems shows
they give consistent results.

\begin{table*} 
\caption{ Table showing results of rank Spearman correlation tests
between the measured line width and the luminosity of the source for
the $M-z$ composite spectra. The number of data points, $N$, and the rank
Spearman statistics, $\rho$, are shown together with the probability
of that value of $\rho$ being obtained by chance. These statistics are
shown for the FWHM and IPV methods of measuring the velocity
widths of the emission line and for the data before and after
contaminating emission was subtracted.}
\label{table_rho}
\begin{center}
\begin{tabular}{lrrrrrr}
\hline
(a)FWHM Method& \multicolumn{3} {c}{Entire Feature} & \multicolumn{3}{c}{Contamination subtracted }  \\
Line     & $N$  & $\rho$ &  $P$  & $N$  & $\rho$ &  $P$ \\
\hline
    Ly$\alpha$ & 19  & 0.789    &   5.84e-05  &  18  & 0.564   &  1.47e-02 \\  
     SiIV      & 25  & 0.124    &   5.55e-01  &  25  & 0.124   &  5.55e-01 \\  
    CIV        & 34  & 0.669    &   1.50e-05  &  34  & 0.679   &  1.03e-05 \\  
CIII]+SiIII]   & 48  & 0.618    &   2.88e-06  &  48  & 0.426   &  2.55e-03 \\ 
     MgII      & 41  & 0.574    &   2.82e-04  &  41  & 0.574   &  8.65e-05 \\  
    $[$NeV]    & 23  & 0.375    &   7.83e-02  &  23  & 0.375   &  7.83e-02 \\  
    $[$OII]    & 31  & 0.139    &   4.56e-01  &  31  & 0.139   &  4.56e-01 \\  
  $[$NeIII]    & 24  & 0.258    &   2.23e-01  &  24  & 0.283   &  1.81e-01 \\  
    \hd        & 14  & 0.473    &   8.80e-02  &  14  & 0.240   &  4.09e-01 \\  
     \hg       & 20  & 0.606    &   4.62e-03  &  20  & 0.641   &  2.34e-03 \\  
     \hb       & 15  & 0.775    &   6.90e-04  &  15  & 0.836   &  1.04e-04 \\  
    $[$OIII]   & 17  & 0.642    &   5.45e-03  &  17  & 0.642   &  5.45e-03 \\ 
\hline
(b) IPV Method& \multicolumn{3} {c} {Entire Feature} & \multicolumn {3}{c} {Contamination subtracted} \\
Line          & N  & $\rho$ &  P & N  & $\rho$ & P \\
\hline
    Ly$\alpha$& 18  &   0.740  &  4.47e-04  &  17  &   0.716  &  1.24e-03\\
    SiIV      & 30  &$-$0.054  &  7.77e-01  &  30  &$-$0.054  &  7.77e-01\\
   CIV        & 34  &   0.678  &  1.05e-05  &  34  &   0.623  &  8.42e-05\\
CIII]+SiIII]  & 48  &   0.601  &  6.35e-06  &  48  &   0.350  &  1.47e-02\\
    MgII      & 43  &   0.490  &  8.41e-04  &  43  &   0.408  &  6.61e-03\\
   $[$NeV]    & 24  &   0.712  &  9.37e-05  &  24  &   0.712  &  9.47e-05\\
   $[$OII]    & 31  &   0.167  &  3.68e-01  &  31  &   0.167  &  3.68e-01\\
 $[$NeIII]    & 25  &   0.477  &  1.59e-02  &  25  &   0.481  &  1.50e-02\\
   \hd        & 23  &   0.447  &  3.26e-02  &  23  &   0.242  &  2.66e-01\\
    \hg       & 23  &   0.752  &  3.51e-05  &  23  &   0.611  &  1.97e-03\\
    \hb       & 17  &   0.400  &  1.12e-01  &  17  &   0.652  &  4.57e-03\\
   $[$OIII]   & 15  &   0.907  &  3.06e-06  &  16  &   0.906  &  1.36e-06\\
\hline
\end{tabular}
\end{center}
\end{table*}

\subsubsection{FWHM}

The FWHM is defined as the width of the line at the position where the
line flux falls to half its peak value. The peak flux was taken to be
the maximum flux within $l_{c}$ $\pm$ 0.75$f_{G}$, where $l_{c}$ and
$f_{G}$ are the central wavelength and full width at half maximum
obtained from the Gaussian fit to the data. In the case of lines
fitted with multiple Gaussians $f_{G}$ was taken to be that of the
broadest component. Interpolation between adjacent 1\AA\ bins was used
to estimate the FWHM to a fraction of an Angstrom and in the few cases
where the line flux fell below half of the maximum flux and then rose
back above it, the largest possible width was used. This method has
the advantage that unlike a Gaussian fit to the emission line, no
assumptions are made about the shape of the line profile but in low
S/N data it is possible for the width of the line to be
overestimated. The error in the FWHM is estimated from the error in
the flux at the half-maximum positions and converted to an error in
wavelengths using the gradient of the line profile at the half-maximum
position. If the flux in the line fell below zero at any point between
$l_{c}$ $\pm$ 0.75$f_{G}$ the velocity width of that line was not
measured.

\subsubsection{IPV}

Here we use the integrated flux of the emission line to obtain an
estimate of the line width. The cummulative flux in the line was found
as a function of wavelength, beginning at the far blue side of the
line ($l_{c}- 1.5\times f_{G}$) with a cummulative flux = 0 and ending
at the far red side of the line ($l_{c}+ 1.5\times f_{G}$), where the
cummulative flux is equal to the total integrated flux of the line
($F$).  The wavelengths at which the cummulative flux = 0.25$F$ and
0.75$F$ were found and their difference provided an estimate of the
line width. Like the FWHM method described above, this method does not
rely on any assumptions about the line profile and has the added
advantage that all the flux in the line is used rather than just the
core of the emission line. The error in the line width obtained via
this method was estimated from the errors in the total integrated flux
and the integrated flux at the 25 per cent and 75 per cent
positions. The error in flux was then converted to an error in
angstroms using the local gradient of the cummulative flux at that
position. Since this method relies on the integrated line flux (rather
than the peak flux) it can be used to measure widths for spectra with
a lower SNR and we therefore obtained more line width measurements
using this method than for the median width.

\section{Results}

\begin{table*} 
\caption{ Table showing results of partial Spearman correlation tests
between the measured line width, the redshift and the luminosity of
the sources contributing to the $M-z$ composite spectra, taking into
account the correlation between redshift and luminosity. We show the
partial rank Spearman correlation, $\rho$, and the probability of that
value of $\rho$ being obtained by chance. The number of data points
used in this analysis, $N$, is the same as that listed in Table 1.  The
statistics are shown for the FWHM and IPV methods of measuring the
velocity widths of the emission line and for the data before and after
contaminating emission was subtracted.}
\label{table_partialrho}
\begin{center}
\begin{tabular}{lrrrrrrrr}
\hline
(a) FWHM Method& \multicolumn{4} {c}{Entire Feature} & \multicolumn{4}{c}{Contamination subtracted }  \\
Line           &$\rho_{v-z}$ &  $P_{v-z}$ & $\rho_{v-L}$ &  $P_{v-L}$& $\rho_{v-z}$ &  $P_{v-z}$ & $\rho_{v-L}$ &  $P_{v-L}$ \\
\hline
    Ly$\alpha$  & $-$0.071 &   7.72e-01&  0.756 &  1.79e-04   &   0.092 & 7.17e-01 &  0.508 &   3.12e-02 \\  
     SiIV       &$-$0.534  &   5.94e-03&  0.383 &  5.91e-02   &$-$0.534 & 5.94e-03 &  0.383 &   5.91e-02 \\  
    CIV         &$-$0.674  &   1.23e-05&  0.835 &  8.55e-10   &$-$0.665 & 1.77e-05 &  0.835 &   8.13e-10  \\  
 CIII$]+$SiIII$]$&$-$0.271 &   6.25e-02&  0.630 &  1.58e-06   &$-$0.360 & 1.21e-02 &  0.536 &   8.72e-05  \\ 
     MgII       &$-$0.036  &   8.24e-01&  0.375 &  1.57e-02   &$-$0.036 & 8.24e-01 &  0.375 &   1.57e-02  \\  
    $[$NeV]     &$-$0.116  &   5.98e-01&  0.270 &  2.12e-01   &$-$0.116 & 5.98e-01 &  0.270 &   2.12e-01  \\  
    $[$OII]     &$-$0.269  &   1.43e-01&  0.300 &  1.02e-01   &$-$0.269 & 1.43e-01 &  0.300 &   1.02e-01  \\  
  $[$NeIII]     &$-$0.426  &   3.81e-02&  0.486 &  1.62e-02   &$-$0.374 & 7.20e-02 &  0.453 &   2.61e-02  \\  
    \hd         &$-$0.005  &   9.86e-01&  0.174 &  5.52e-01   &  0.196  & 5.02e-01 & $-$0.108 & 7.13e-01  \\  
     \hg        &  0.029   &   9.03e-01&  0.257 &  2.74e-01   &$-$0.088 & 7.11e-01 &  0.374 &   1.04e-01  \\  
     \hb        &  0.161   &   5.66e-01&  0.479 &  7.07e-02   &  0.893  & 7.25e-06 &  0.195 &   4.87e-01  \\  
    $[$OIII]    &$-$0.442  &   7.56e-02&  0.710 &  1.39e-03   &$-$0.442 & 7.56e-02 &  0.710 &   1.39e-03  \\ 
\hline
(b) IPV Method& \multicolumn{4} {c} {Entire Feature} & \multicolumn {4}{c} {Contamination subtracted} \\
Line          & $\rho_{v-z}$ & P$_{v-z}$ & $\rho_{v-L}$ & P$_{v-L}$& $\rho_{v-z}$ &  P$_{v-z}$ & $\rho_{v-L}$ & P$_{v-L}$ \\
\hline
    Ly$\alpha$&  0.425  &   7.86e-02  &  0.507  & 3.45e-03 &  0.725   & 9.88e-04 &  0.707  &  1.49e-03  \\
    SiIV      &$-$0.390 &   3.33e-02  &  0.137  & 4.70e-01 & $-$0.390 & 3.33e-02 &  0.137  &  4.70e-01 \\
   CIV        &$-$0.479 &   4.19e-03  &  0.764  & 1.51e-07 & $-$0.475 & 4.49e-03 &  0.725  &  1.22e-06 \\
CIII$]+$SiIII$]$&$-$0.153&  2.99e-01  &  0.567  & 2.63e-05 & $-$0.305 & 3.54e-02 &  0.452  &  1.28e-03 \\
    MgII      &$-$0.652 &   2.05e-06  &  0.750  & 6.93e-09 & $-$0.682 & 4.87e-07 &  0.744  &  1.10e-08 \\
   $[$NeV]    &   0.282 &   1.83e-01  &  0.241  & 2.58e-01 &  0.282   & 1.83e-01 &  0.241  &  2.58e-01  \\
   $[$OII]    &$-$0.361 &   4.61e-02  &  0.391  & 2.97e-02 & $-$0.361 & 4.61e-02 &  0.391  &  2.97e-02 \\
 $[$NeIII]    &$-$0.287 &   1.64e-01  &  0.472  & 1.73e-02 & $-$0.161 & 4.43e-01 &  0.379  &  6.16e-02 \\
   \hd        &   0.298 &   1.67e-01  &$-$0.023 & 9.17e-01 &  0.355   & 9.68e-02 &$-$0.201 &  3.59e-01 \\
    \hg       &   0.041 &   8.52e-01  &  0.488  & 1.81e-02 & $-$0.157 & 4.74e-01 &  0.477  &  2.14e-02 \\
    \hb       &   0.343 &   1.77e-01  &  0.705  & 7.88e-01 &  0.010   & 9.69e-01 &  0.516  &  3.41e-02 \\
   $[$OIII]   &$-$0.534 &   3.87e-02  &  0.903  & 4.05e-06 & $-$0.664 & 5.05e-03 &  0.931  &  1.66e-07 \\
\hline
\end{tabular}
\end{center}
\end{table*}

\subsection{Dependence of line velocity widths on QSO luminosity}

We derived the Spearman rank correlation coefficients for the
correlation of line width (as measured by each method) with the source
luminosity for each line (Table 1a and b). The emission lines widths
were corrected for the instrumental resolution ($\delta \lambda$=9\AA\
for $\lambda$=5500\AA\ in the observed spectrum) by assuming that both
the emission line and the instrumental resolution were Gaussians. The
small number of measurements with widths narrower than the
instrumental resolution were excluded as were velocity measurements
with fractional errors of more than 20 per cent. We measured the
correlation of the line width both before and after the contaminating
emission was subtracted in order to determine whether our fitting
procedure could be introducing spurious correlations into the data.
 
{\it A priori} we chose a significant correlation to be one with a
formal probability $P < 1$ per cent of being due to chance. For the
$M-z$ spectra significant correlations of line width with luminosity
were found for the broad emission lines \civ, \ciii, \mgii\ and the
Balmer lines \hg\ and \hb\ (Table 1). With the exception of the \hb\
emission line, these correlations were independent of the method used
to measure the line width and whether the contaminating emission had
been subtracted. For \hb\ only the IPV  of the whole feature (\hb\
plus the two \oiii\ lines) failed to show a significant correlation
between luminosity and line width, most likely due to the contribution
of the \oiii\ lines to the red side of the line.

A highly significant correlation ($P <0.1$ per cent) was also found
between the width of the combined \lya\+\nv\ feature and the source
luminosity but not in the \nv-subtracted \lya\ line. This probably
reflects the difficulties of accurately fitting the components in the
\lya\+\nv\ feature rather than a lack of a correlation between the
width of \lya\ and luminosity. The blue side of \lya\ emission line is
subject to a considerable amount of absorption and it is probable that
\lya\ is not well modeled using a combination of symmetrical line
profiles, such as a Gaussians. Any errors in the fit to the \lya\
component would also cause errors in the fit to the \nv\ component and
the \nv-subtracted \lya\ is therefore probably not a good
representation of the true line. Additionally, any variation in the
relative strength of the \lya\ and \nv\ lines, for example if \nv\
became relatively stronger with luminosity, would also alter the FWHM
of the feature.

Aside from \lya, the only broad lines which do not exhibit a
significant correlation between line width and source luminosity in
the $M-z$ spectra are \siiv\ and \hd. \hd\ is a relatively weak
emission line and it is difficult to  measure its width accurately. On
the other hand \siiv\ does not appear to vary significantly at all,
either in flux or velocity (Croom et al. 2002).

Of the narrow emission lines, \oiii\ is the only line which exhibits a
significant correlation with luminosity (Table 1).

\subsection{Dependence of line widths on redshift}

\begin{figure*}
\centering
\centerline{\psfig{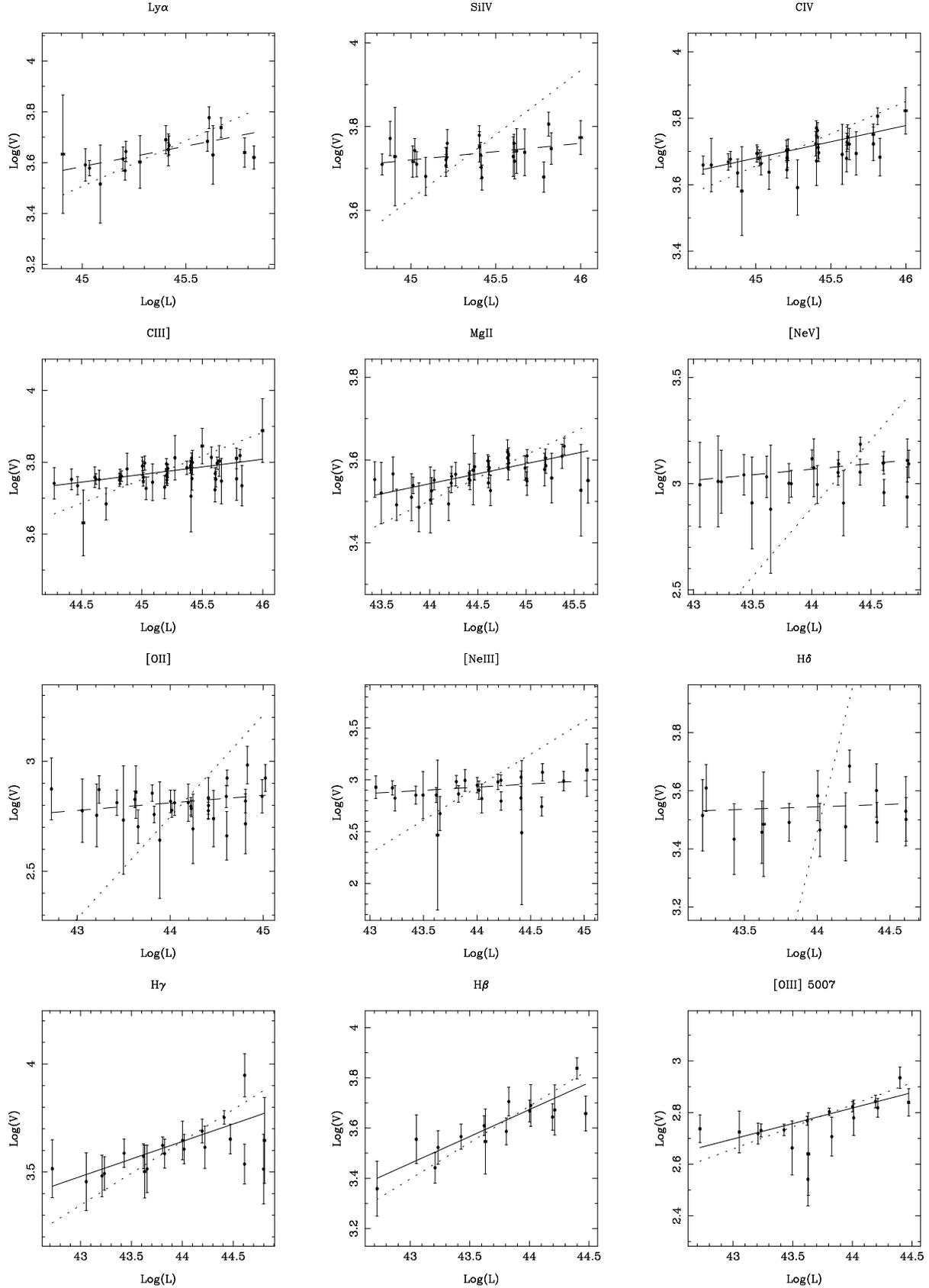}}
\caption{Velocity ($v$)-luminosity (L$_{b}$) plots for the FWHM of the
contamination subtracted spectra. Velocity is in units of \kms\ and
$\lb$ in units of erg s$^{-1}$. The best fits of the form
$\log(v)=a+b\log(L_{b})$ are also shown for each line (solid  line for
data which showed a significant correlation between velocity and
luminosity; dashed lines for data which showed no significant
correlation). We also show the best fit derived from regressing
luminosity upon velocity ($\log(L_{b})=c+d\log(v); $dotted line).The
errors bars are scaled such that $\chi^{2}$ is equal to the number or
degrees of freedom.}
\label{vel_bm_m}
\end{figure*}

\begin{table*} 
\caption{ Table showing the best fits for the velocity width, $v$ (km s$^{-1}$),
correlation with luminosity, L$_{b}$ (erg s$^{-1}$). A linear fit of the form
$\log(v)=a+b\log(\rm{L}_{b})$ was applied to the data.}
\label{table_lslope}
\begin{center}
\begin{tabular} {lrrrrrrrr}
(a) Median FWHM & \multicolumn{4} {c} {Entire feature} & \multicolumn{4}{c} {Contamination subtracted}  \\
Line               &   $a$  &$\sigma_{a}$& $b$  &$\sigma_{b}$& $a$ & $\sigma_{a}$ & $b$ & $\sigma_{b}$ \\ 
\hline   
 Ly$\alpha$    &$-$19.43&  4.50 &  0.511 &  0.099 &$-$3.64 & 2.13 &  0.161 & 0.047  \\
    SiIV       &   2.08 &  1.06 &  0.036 &  0.023 &  2.08 & 1.06  &  0.036 & 0.023  \\
   CIV         &$-$0.78 &  1.01 &  0.099 &  0.022 &$-$0.72 & 0.95 &  0.098 & 0.021  \\
CIII$]+$SiIII$]$&  0.39 &  0.61 &  0.075 &  0.014 &  1.86 & 1.58  &  0.042 & 0.013  \\ 
 MgII          &   1.43 &  0.40 &  0.048 &  0.009 &  1.41 & 0.40  &  0.048 & 0.009  \\  
   $[$NeV]     &   0.75 &  1.48 &  0.053 &  0.034 &  0.75 & 1.48  &  0.053 & 0.034  \\
   $[$OII]     &   1.30 &  1.06 &  0.034 &  0.024 &  1.30 & 1.06  &  0.034 & 0.024  \\
   $[$NeIII]   &$-$1.96 &  3.08 &  0.112 &  0.070 &  0.24 & 2.97  &  0.061 & 0.067  \\
   H$\delta$   &   0.77 &  1.30 &  0.062 &  0.029 &  2.79 & 1.93  &  0.017 & 0.044  \\ 
  H$\gamma$    &$-$0.83 &  1.72 &  0.099 &  0.039 &$-$3.49 & 1.77 &  0.162 & 0.040  \\
  H$\beta$     &$-$4.03 &  1.66 &  0.169 &  0.038 &$-$5.75 & 1.41 &  0.214 & 0.032  \\  
$[$OIII$]$ 5007&$-$2.48 &  1.87 &  0.120 &  0.043 &$-$2.45 & 1.89 &  0.120 & 0.043  \\ 
\hline
(b)IPV       &\multicolumn{4}{c}{Entire Feature} & \multicolumn{4}{c}{Contamination Subtracted}\\
Line               & $a$ &$\sigma_{a}$ &  $b$  & $\sigma_{b}$ &  $a$ &$\sigma_{a}$ & $b$ & $\sigma_{b}$ \\   
\hline 
 Ly$\alpha$    &  1.62  &  0.90 &  0.048 & 0.020 & $-$1.93 & 2.34 & 0.123 & 0.051  \\
    SiIV       &  2.84  &  0.86 &  0.016 & 0.019 &   2.84  & 0.86 & 0.016 & 0.019  \\
   CIV         &  0.92  &  0.67 &  0.063 & 0.015 &   0.40  & 0.77 & 0.072 & 0.017  \\
CIII$]]+$SiIII$]$ & 1.09 & 0.58 &  0.057 & 0.013 &   2.93  & 0.36 & 0.015 & 0.011  \\ 
 MgII          &$-$0.59 &  1.55 &  0.099 & 0.035 &   1.34  & 1.01 & 0.053 & 0.023  \\  
   $[$NeV]     &$-$1.92 &  0.90 &  0.110 & 0.020 & $-$1.92 & 0.89 & 0.110 & 0.020  \\
   $[$OII]     &$-$1.03 &  0.94 &  0.036 & 0.021 &   1.03  & 0.92 & 0.036 & 0.021  \\
   $[$NeIII]   &$-$5.48 &  2.41 &  0.188 & 0.054 &$-$1.44  & 1.99 & 0.095 & 0.045  \\
   H$\delta$   &  1.57  &  1.87 &  0.041 & 0.042 &   2.58  & 1.49 & 0.018 & 0.034  \\ 
  H$\gamma$    &$-$3.16 &  1.11 &  0.149 & 0.025 &$-$3.42  & 1.50 & 0.156 & 0.034  \\
  H$\beta$     &$-$4.81 &  3.05 &  0.193 & 0.069 &   0.16  & 0.95 & 0.076 & 0.022  \\  
$[$OIII$]$ 5007&$-$7.44 &  1.31 &  0.231 & 0.030 &$-$6.77  & 1.27 & 0.216 & 0.029  \\ 
\hline
\end{tabular}
\end{center}
\end{table*}

To investigate whether there is any evidence for evolution in the AGN
line width (i.e. narrower or broader lines observed at higher
redshifts), we derived the Spearman rank correlation coefficients for
a correlation of line width with redshift. Significant positive
correlations between redshift and velocity were found for some methods
of measuring the velocity of the emission lines (e.g. the FWHM
velocity of \mgii, \hg, and \hb; the IPV of \nev) but unlike the
luminosity correlations, no line showed a significant correlation
between redshift and velocity for every method.

A rank Spearman correlation test of our data reveals highly
significant correlations ($P<0.01$; generally $P<10^{-5}$ ) between
luminosity and redshift for all the emission lines except \siiv\ and
\lya\ and it is entirely possible that the handful of significant
correlations we find between redshift and velocity are due to a
combination of the redshift-luminosity ($z-L$) correlation and that
seen between luminosity and velocity ($v-L$ correlation). We therefore
derived the partial Spearman correlation coefficients (see Croom et
al. 2002 for a detailed discussion of this analysis method) for the
data. In this analysis we test the hypothesis that any observed
correlation between velocity and redshift is due to the $z-L$
correlation and the $v-L$ correlation with the coefficients and their
associated probabilities shown in Table 2. For comparison we repeated
this analysis for the $v-L$ correlation, testing the hypothesis that
the $v-L$ correlation is due to a $v-z$ correlation and the $z-L$
correlation (again shown in Table 2). We find that while the velocity
is clearly correlated most strongly with luminosity for the \civ,
\ciii\ and \mgii\ emission lines, there is some evidence for a weak
anti-correlation between velocity and redshift, significant only for
the \civ\ emission line. If there is a weak trend for the emission
lines to become narrower as redshift increases, this could easily be
masked by the strong positive $v-L$ and $L-z$ correlations, particularly
as the range of redshifts over which the different emission lines are
observed is limited. The possible dependence of line velocity with
redshift is discussed further in Section 5.2. None of the narrow
emission lines exhibit any evidence for a correlation between redshift
and velocity.

We note that neither the \hb\ or \hg\ lines exhibit a significant $v-L$
correlation in the partial Spearman correlation analysis although both
lines show strong positive correlations using the standard Spearman
correlation. They are also the only lines which exhibit a weak
positive $v-z$ correlation. This may be because the they are not
observed over a sufficiently large redshift range for the effects of
the $z-L$, $v-L$ and $v-z$ correlations to be disentangled using this
analysis (\hb, for example, is only observed in three redshift
bins). In any case it is clear from Figure 2 and the regression line
analysis (Section 4.3) that both line widths of both \hb\ and \hg\ are
correlated positively with luminosity.

The interpretation of the partial Spearman correlation analysis for
the \mgii\ emission line is also complex. Our results indicate the
\mgii\ line exhibits a weak correlation between the FWHM velocity
width and luminosity and a much stronger correlation between the IPV
width and luminosity. The IPV widths also show a significant
correlation with redshift. We have been unable to correct for the
\feii\ emission underlying the \mgii\ line and if the relative
strength of the \feii\ features is a function of luminosity or
redshift, this could cause the behaviour seen below in the \mgii\
line. For example, an increase in QSO metallicity with redshift would
increase the strength of the \feii\ emission relative to that of the
\mgii\ line and hence increase the width of the broad base (see Figure
1 and Section 3.1). 

\subsection{Fits to the velocity-luminosity relationship }

For our composite spectra, the $\bj$ luminosity L$_{b}$ was estimated
from the mean absolute $\bj$ magnitude of the spectra contributing to
the emission line in each composite. The best fit to the relationship
$\log(v)=a + b\log(L_{b})$ was obtained from a standard
least-squares fit to the velocity width of the emission lines and QSO
luminosity dependence. The parameters $a$ and $b$ fitted to each line
are given in Table 3 and a representative data-set is shown in Figure
2 (the FWHM velocity width of the contamination subtracted data). It
is likely that the errors in the velocity widths have been
underestimated in our analysis due to  the method used to normalize
the spectra, underlying \feii\ features errors in the subtraction of
the contaminating emission. The estimated errors on the emission line
widths do not take these factors into account and hence the errors on
$a$ and $b$ derived using the standard method are probably an
underestimate of the true error on the fit. We therefore used
$\chi^{2}$ from the fit to rescale the errors on the velocity so that
$\chi^{2}$ is equal to the number of degrees of freedom and hence
achieve a more realistic estimate of the errors on $a$ and $b$.

As a further check of the correlation analysis performed above, we
refit the data, this time fitting a line of the form
$\log(\rm{L}_{b})=c+d\log(v)$. For a perfect correlation between
velocity width and luminosity (and no errors in either parameter) the
best fit lines of $\log(\rm{L}_{b})$ upon $\log(v)$ will coincide with those
of $\log(v)$ upon $\log(\rm{L}_{b})$. For no correlation between $v$ and
L$_{b}$ the lines of best fit will be perpendicular.

The FWHM of the broad hydrogen lines, \hb, \hg\ and \lya\ show a
slightly steeper dependence on luminosity than the other emission
lines (Table 3).  The best fit slopes to \mgii, \civ\ and \siiv\ lie
in the range $0.04<b<0.1$ whereas the slope of the hydrogen lines  is
in the range $0.1<b<0.2$.

Previous investigations into a possible correlation between velocity
width and QSO luminosity have produced ambiguous results. From
observations of the \hb\ emission line in individual sources K00
report an anticorrelation between velocity and luminosity with a slope
(in log-log space) of -0.27.  On the other hand, several authors
(e.g. Wandel \& Yahil 1985; Joly et al. 1985; Boroson \& Green 1992;
Stirpe, Robinson \& Axon 1999)
\nocite{Wandel85}\nocite{Joly85}\nocite{Stirpe99} have found the width
of the Balmer lines (either FWHM, full width at zero intensity or full
widths at some percentile of the maximum flux) to be weakly correlated
luminosity. These studies were all based on samples containing $<$150
sources and it is possible that the inconclusive results may be due to
the intrinsic scatter in the $v-L$ relationship as well as selection
effects.

\section{Discussion}
\subsection{Broad emission lines and black hole mass estimates} 

\begin{figure*}
\centering 
\psfig{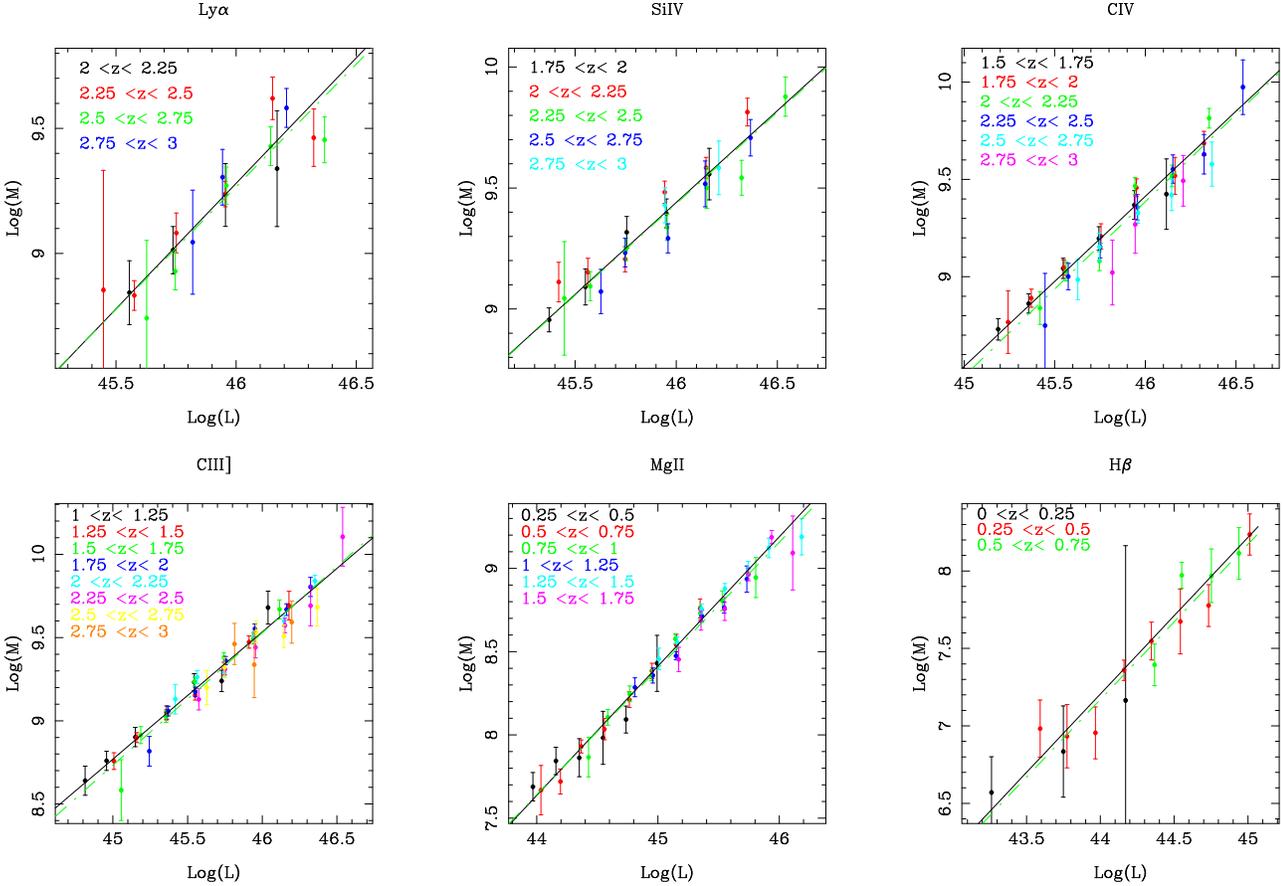}
\caption{The mass ($M$)--luminosity ($L_{5100}$) relationship for the
different broad emission lines. Mass is given in units of $M_{\sun}$
and luminosity in erg s$^{-1}$. Different colours reflect the
different redshift ranges as shown in the key for each plot and the
error bars shown have been scaled such that $\chi^{2}$ is equal to the
number of degrees of freedom. The YonX (solid black line) and BCES
(green dot-dash line) fits to the $\mbh-L$ relationship for each line are
also shown.}
\label{vel_bm_noscal}
\end{figure*}

In reverberation mapping studies (e.g. Wandel et al. 1999, Onken \&
Peterson 2002)\nocite{Wandel99}\nocite{Onken02} the rms spectrum of
the \hb\ emission line is often used to estimate the velocity
dispersion of the gas. The rms spectra, formed from the rms difference
between the individual spectra obtained during the monitoring
programme and the mean spectrum, measures only the components which
vary and, since the emission from the NLR does not vary significantly
on short time scales it should not be present in the rms spectrum. At
first sight it therefore seems reasonable that the velocity dispersion
of the BLR gas would be best represented by the \hb\ velocity width
measured from the line after subtraction of the narrow emission line
component. However, K00 found that there was little difference between
the \hb\ line width obtained from the rms spectrum and that measured from
the mean spectrum without correcting for the narrow line
emission. Vestergaard (2002) also found that the width of the \hb\
line in the rms spectrum was more similar to that of a single-epoch
observation before any correction for NLR emission. This implies that
much of the narrow core of the \hb\ emission line (Figure 1) is from
gas associated with the BLR (rather than the NLR) and we therefore use
the \hb\ line width from the composite spectra before correction for any
narrow line emission in the analysis which follows. For the non-Balmer
lines, we use the FWHM line widths measured after contaminating
emission has been subtracted.
  
A secondary issue is the underlying \feii\ emission.  As discussed in
Section 3.1, many of the \feii\ emission features have been treated as
continuum, particularly the large feature underlying \mgii, and it is
therefore not possible for us to correct the emission lines for the
underlying \feii. This may impact on our measurement of the \mgii\
FWHM. In their study, McLure \& Jarvis (2002) used \mgii\ FWHM
measurements from spectra taken with the International Ultraviolet
Explorer (IUE) which had been corrected for the underlying \feii\
emission. They found a linear almost one-to-one, correlation between
the FWHM measured for the \hb\ and \mgii\ lines whereas we find that
the \mgii\ emission is generally broader than the \hb\ emission. On
the other hand, some of the \hb\ FWHM measurements used by McLure \&
Jarvis (2002) have been corrected for narrow \hb\ emission and in
these cases it is reasonable to asume that their \hb\ FWHM would be
broader than ours.

There is also an \feii\ emission line to the red of \hb\ ($\sim$ 4930
\AA, see e.g. Vanden Berk et al. 2002) which we have not removed
before measuring the \hb\ FWHM.  Vestergaard (2002) found that it was
not necessary to correct for the FeII emission unless the FeII
emission from the object was unusually strong, and as there is no
evidence for strong \feii\ emission in our composite spectra, we do
not believe this feature will have a significant impact on the \hb\
FWHM.  In any case, we note that K00 did not correct for \feii\
emission when measuring the \hb\ FWHM from their mean spectra.

Having determined which measurements of the line widths most
accurately represent the velocity dispersion of the gas, we need to
obtain an estimate of the radius of the BLR.  From reverberation
mapping experiments, K00 found the BLR radius, $r\propto$
$L^{0.7}_{5100}$ where $L_{5100}$ is the continuum luminosity at
5100\AA\ (defined as $\lambda L_{5100}$ in K00 and $=3.56\rm{L}_{b}$ for
our data). This result depends on the cosmology assumed and the
analytic method used to determine the fit. K00 assumed $q_0=0.5$,
$\lo=0$ and $H_0=75$~\kms~Mpc$^{-1}$ and used the {\it fitexy} linear
regression method (Press et al. 1992), which takes into account errors
in both $L$ and $r$, to fit their data.  Re-analysing the K00 data,
assuming the cosmology adopted here ($\Om=0.3$, $\lo=0.7$ and
$H_0=70$~\kms~Mpc$^{-1}$), Netzer (2003) finds a similar relationship
with
\begin{equation}
r=27.4  (\frac{L_{5100}} {10^{44}{\rm erg~s^{-1}}}) ^{0.68\pm0.03}~~
{\rm light~days}.
\label{eq:rl}
\end{equation}

Using the BCES bisector method (Akritas \& Bershady 1996) to fit the
same data, Netzer (2003) obtains a somewhat flatter relationship,
$r\propto L_{5100}^{0.58\pm0.12}$, although the two fits are formally
consistent. Similar results are obtained by McLure \& Jarvis (2002)
who find $r\propto L_{5100}^{0.61\pm0.10}$ for our cosmology and the
BCES bisector method. Here we have adopted $r\propto L_{5100}^{0.68}$
(Equation \ref{eq:rl}) but we note that a flatter $r--L_{5100}$
relationship will yield smaller black hole masses at higher
luminosities.

The black hole mass is then given by (K00):
\begin{equation}
\mbh=1.456 \times 10^{5} (\frac{r} {\rm{light~days}}) (\frac{v}{1000 {\rm km
~s^{-1}}})^{2} M_{\sun}
\label{eq:bhmass}
\end{equation}

where $v$ is the FWHM of the \hb\ emission line. The BLR clouds are
assumed to be in random orbits about the black hole and thus in
Equation \ref{eq:bhmass} the velocity dispersion is assumed to be
$(\sqrt 3 /2) v$ (Netzer 1990). If the orbits of the BLR clouds are not
random but confined to disks, as suggested by e.g., Young et al. 1999,
McLure \& Dunlop (2002), the velocity dispersion is $(3/2) v$, and the
black hole masses are a factor of 3 larger than we derive.

\begin{figure}
\centering \centerline{\psfig{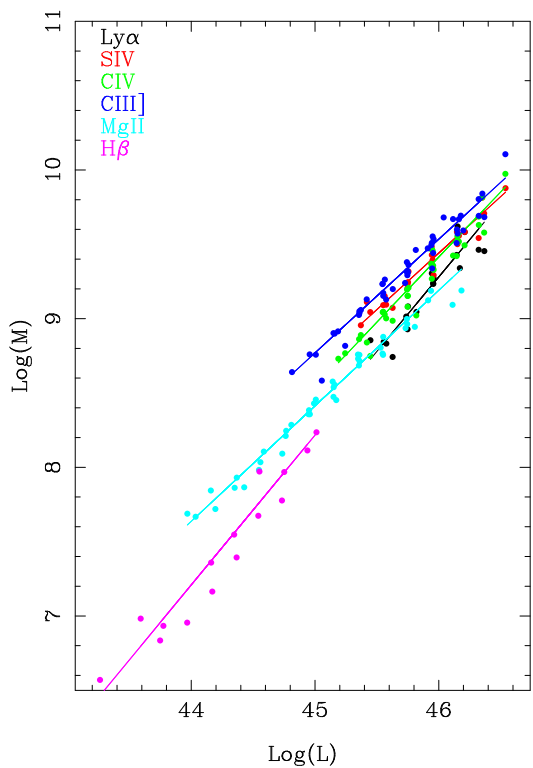}}
\caption{The mass ($M$)--luminosity ($L_{5100}$) relationship for
selected broad emission lines, each line shown in a different
colour. The best BCES fit to the $\mbh-L$ relationship for each line is
also shown. Error bars are not plotted for clarity.}
\label{fig_allmass}
\end{figure}

The black hole masses for selected lines were estimated using
Equations \ref{eq:rl} and \ref{eq:bhmass} and are shown in Figure
\ref{vel_bm_noscal} plotted  against $L_{5100}$. As expected, a strong
correlation can be seen between BH mass ($M$) and the luminosity,
$L_{5100}$, for each line with remarkably little scatter, generally
less than 0.3 dex. The best fit of the form $\log(M)=a +\beta \log(L)$
for each line is shown in Figures \ref{vel_bm_noscal} and
\ref{fig_allmass} with the parameters for each fit given in Table
\ref{table_mslope}. The data were fitted using both the standard
linear regression method (YonX) and the {\it BCES} method. From Table
\ref{table_mslope} it is clear that the two methods agree to within
the errors (see also Figure \ref{vel_bm_noscal}). For the linear
regression method the errors in the fit parameters were evaluated as
described in the previous section, using the $\chi^{2}$ of the fit to
derive the errors in the mass estimate and hence $a$ and $\beta$.  We
note that the linear relationship between $\mbh$ and $L_{5100}$
derived for the \hb\ line is similar to that obtained for the \lya\
emission line in this analysis. Fits to the other emission lines are
slightly offset from, and flatter than, the fit to the hydrogen lines
(Figure \ref{fig_allmass}). Nevertheless, for a given continuum
luminosity, the black hole masses obtained from the different emission
lines typically lie  within $\pm0.3$ dex of each other.

\begin{table*} 
\caption{The best fits to the relationship between the black hole
mass, $M$, with the monochromatic luminosity, $L_{5100}$, in units of
erg s$^{-1}$. Black hole mass is in units of $M_{\sun}$. A linear fit
of the form  $log(M)=a + \beta log(L_{5100})$ was applied to the data,
where $\beta$ is the slope of the relationship and $a$ is a
constant. $\delta$ is the normalisation constant applied to the data
to force the weighted mean of the line to lie on the \hb\ YonX
relationship and has errors of $\sim$ 0.1dex }
\label{table_mslope}
\begin{center}
\begin{tabular} {lccccccr}
\hline
Line   & method& $a$ & $\sigma_{a}$ & $\beta$ & $\sigma_{\beta}$ & $\delta$ \\
\hline
\lya\  &YonX&$-$36.79 &  4.32 &  1.00 &  0.09 & $-$0.06 \\
       &BCES&$-$35.82 &  5.08 &  0.98 &  0.11 &          \\    
\siiv\ &YonX&$-$25.49 &  1.82 &  0.76 &  0.04 &$-$0.24 \\ 
       &BCES&$-$25.38 &  1.75 &  1.02 &  0.12 &        \\     
\civ\  &YonX&$-$30.85 &  1.93 &  0.87 &  0.04 &$-$0.22 \\
       &BCES&$-$31.90 &  1.86 &  0.90 &  0.04 &        \\       
\ciii\ &YonX&$-$25.65 &  1.17 &  0.76 &  0.03 &$-$0.38 \\ 
       &BCES&$-$27.09 &  1.45 &  0.80 &  0.03 &        \\      
\mgii\ &YonX&$-$26.54 &  0.80 &  0.78 &  0.02 &$-$0.15 \\ 
       &BCES&$-$25.71 &  0.88 &  0.76 &  0.02 &        \\            
\hb\   &YonX& $-$37.12&  3.32 &  1.01 &  0.08 & 0.0 \\    
       &BCES&$-$36.90 &  2.81 &  1.00 &  0.06 &  \\
all data& YonX&$-$32.05&0.08 & 0.90 & 0.00  \\
        & BCES&$-$33.53&0.78 & 0.93 & 0.02  \\
\hb\ and& YonX&$-$36.68&0.08 & 1.00 & 0.00  \\
\civ\ only& BCES&$-$36.77&0.74&1.00& 0.02  \\     
\hline
\end{tabular}
\end{center}
\end{table*}

Reverberation mapping experiments (e.g., Peterson \& Wandel 1999,
2000; Onken \& Peterson 2002) find that different emission lines
respond to changes in the UV continuum with different time lags. This
indicates that the BLR is stratified, with the broader emission lines
(such as \civ) originating closer to the central engine than the
Balmer lines. The empirical $r-L_{5100}$ relationship (Equation \ref{eq:rl})
was obtained by K00 from the \hb\ and \ha\ emission lines and thus its
use to determine black hole masses via Equation \ref{eq:bhmass} will lead to an
overestimate of the black hole mass for \civ. To obtain more reliable
estimates of $M$ for the UV lines it is necessary to introduce a
``calibration'' factor to equation \ref{eq:rl} to correct for the difference
between the radius of the gas emitting the broad UV lines, such as
\civ, and that emitting the \hb\ lines. This has already been
attempted by Vestergaard (2002) who compared the BH masses estimated
from reverberation mapping of the \hb\ emission line with those
estimated from the \civ\ FWHM in the same object assuming that, for
the \civ\ line,
\begin{equation} 
\mbh =a (\frac {FWHM_{\civ}} {1000 \rm{km~s^{-1}}})^{2} (\frac
{L_{1350}}{10^{44} \rm{erg~s^{-1}}})^{0.7}
\end{equation}
where $L_{1350}$ is the continuum luminosity at 1350\AA\ (i.e. $\lambda$L$_{1350}$ in Vestergaard 2002) and $a$ is a
constant. She found $a= 10^{6.2\pm0.45}$ which implies that the
radius of the \civ\ emission, $r_{\civ} \sim 0.5^{+1.1}_{-0.4}
r_{H\beta}$ (assuming that the QSO spectra have a spectral index of
$-0.5$) although the errors in $a$, and hence this result, are large.

To determine the calibration factor for our data we re-normalised the
mass-luminosity relationship so that the weighted mean of the mass
derived from each line lies on the \hb\ (YonX) relation, noting that
an extrapolation of the \hb\ mass-luminosity relation to higher
luminosities is entirely consistent with that observed for
\lya. However, because our estimates of black hole mass are based on
the underlying assumption that the K00 radius-luminosity relation
$r\propto L^{0.68}$ holds for all emission lines, we chose not to
adjust the slopes of the observed mass-luminosity relation for each
emission line. The mass-luminosity relationships were therefore
normalised by the addition of a constant offset, $\delta$ (Table
4). Assuming that this offset is due to the emission lines being
emitted from gas situated at different radii, $\delta$ can be
converted to provide an estimate of the radius at which the UV lines
are emitted relative to that of the \hb\ emission (radius =
$10^{\delta}r_{\rm{H}\beta}$). We find that the $r_{\civ} =
0.6^{+0.2}_{-0.1} r_{\rm{H}\beta}$ (Table 4) which is consistent with 
that found by Vestergaard (2002). From their reverberation mapping
study of NGC 3783, Onken \& Peterson (2002) also find that the \civ\
emission line lag is approximately half that of the \hb\ line although
the errors on this measurement are also large.
 
The calibration factor $\delta$ does not address the variation in the
slopes of the mass--luminosity relationships for the different lines,
which, if significant, must result from a difference in the measured
line width-luminosity relation for different species. Since it is
difficult to see how different emission lines could give rise to a
different mass--luminosity relations (a line-independent quantity)
based on the same set of QSO composite spectra, the most
straightforward interpretation of this observation is that the
different slopes are caused by the contaminating emission blended with
the lines. As we have already discussed in this and previous sections,
the \mgii\ and \ciii\ emission lines are blended with emission from
other elements (\feii\ and \siiii\ respectively) which would
effectively broaden the measured FWHM of the lines, making our
calculation of $\delta$ unreliable. The blended emission could also
alter the slope of the $\log(v)-\log(L)$ relation, particularly if
the relative contribution of the lines vary with luminosity or
redshift.  We note that the slope of the $\log(M)-\log(L)$
relationship for \hb\ and \civ\ are statistically identical given
their errors (Table 4) and that these lines also suffer the least
amount of contamination. We cannot, however, rule out the possibility
that the radius-luminosity relation may differ slightly from line to
line.
    
\begin{figure}
\centering \centerline{\psfig{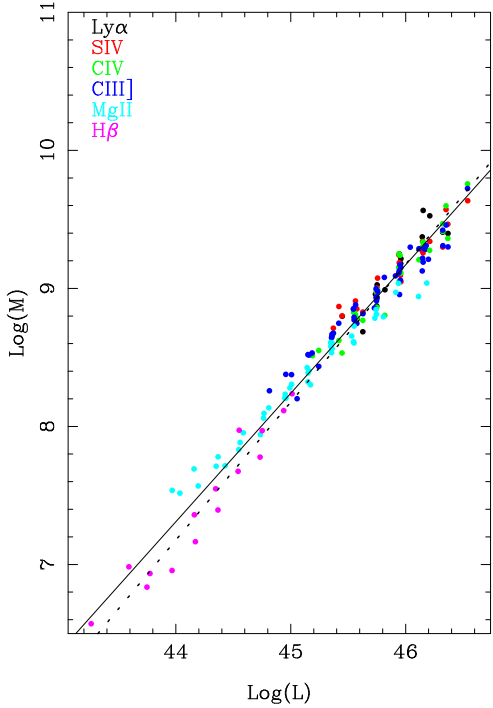}}
\caption{The mass ($M$)--luminosity ($L_{5100}$) relationship for the
different lines normalised to the \hb\ relationship. Each colour
corresponds to a different line. The best BCES fit to the data as a
whole is shown (solid line) as well as that to the \hb\ and \civ\
lines alone (dotted line). Error bars are not plotted in the interests
of clarity. }
\label{fig_allmass_scaled}
\end{figure}

The re-normalised mass-luminosity relations for all lines are plotted
in Figure \ref{fig_allmass_scaled} and we fit an overall
mass-luminosity relation of the form:
\begin{equation}
\log (\mbh) = 0.93(\pm0.05) \log(L_{5100}) - 33.5(\pm2)
\label{eq:ml1}
\end{equation}
to all emission lines (solid line, Fig 5). However, mindful the
problems associated with contaminating emission, we also fit a
mass-luminosity relation to the \hb\ and \civ\ emission lines only
(Fig.5; dotted line) since these lines are the least affected by
contaminating emission and should therefore give the most reliable
estimate. For the \hb\ and \civ\ lines alone, we find

\begin{equation}
\log (\mbh) = 1.00(\pm0.1) \log(L_{5100}) - 36.8(\pm4.5)
\label{eq:ml2}
\end{equation}
which is slightly steeper than that found for all the emission lines
combined.

The errors on the $\mbh-L$ relations are larger than those quoted in
Table 4. In estimating the errors on these relations we have taken
into account the errors on the initial fit to \hb\ which, since it is
used to normalise the mass estimates from the other emission lines,
contributes to the overall errors on the $\mbh-L$ relation. The error
in the fit to the \hb\ and \civ\ lines is larger than that obtained
for all the emission lines as we have fewer data points. Note the
errors on the gradient and the intercept in Equation \ref{eq:ml1}  and
\ref{eq:ml2} are not independent and have a correlation of nearly
-1. The relations we obtain are in reasonable agreement with the
statistical analysis of Netzer (2003), who finds $\mbh\propto
L^{0.9\pm0.15}_{1350}$ for $r_{\rm BLR}\propto L^{0.68}_{1350}$ (where
$L_{1350}$ is the luminosity at 1350\AA) for a sample of over 700
objects (Table 1; Netzer 2003).

An important caveat on the $\mbh-L$ relationship given in equations
\ref{eq:ml1} and \ref{eq:ml2} is that throughout our analysis we have used the relationship
$r_{\rm BLR} \propto L^{\gamma}_{5100}$ discovered by K00 and later
refined by McLure \& Jarvis (2002) and Netzer (2003). As discussed
previously, the value of $\gamma$ is uncertain and depends on the
statistical method used to fit the data, with the {\it fitexy} method
and the BCES method giving values of $0.68\pm0.03$ and $0.58\pm0.12$
respectively for the same data. Although there is a correlation
between velocity and luminosity (Section 4) the slope of the
$\log(v)=a + b\log(L_{5100})$ relationship is quite flat for most of
the emission lines (e.g, $b=0.17$ for \hb, 0.1 for \civ, 0.05 for
\mgii; Table 3) and the slope of the $\log(\mbh)-\log(L_{5100})$
relation is therefore dominated by the value assumed for $\gamma$
which is currently known to an accuracy of only $\pm0.15$. For
example, if we adopt $\gamma=0.58$ in Equation \ref{eq:rl}, we find $\mbh\propto$
$L^{0.83}_{5100}$ for all lines and $\mbh\propto$
$L^{0.9}_{5100}$ for the \hb\ and \civ\ lines alone, again consistent
with Netzer (2003).

Additionally, and perhaps more importantly, it is not known whether
$\gamma$ is constant across all luminosity ranges. The sources in
K00's sample all have $L_{5100} <10^{46}$ erg s$^{-1}$, with only 4
(out of a total sample of 34) having $L_{5100} > 10^{45}$ erg
s$^{-1}$. Our mass estimates for the UV lines are therefore based on
the assumption that the $r_{\rm BLR} \propto L_{5100}$ relation can be
safely extrapolated to higher luminosities. Netzer (2003) found that
eliminating the 3 objects with $L_{5100} < 10^{43}$ erg s$^{-1}$ from
K00's sample resulted in a $\gamma(BCES)=0.71\pm0.21$ and
$\gamma(fitexy)=0.69\pm0.03$ whereas eliminating the seven objects
with $L_{5100} < 10^{43.7}$ erg s$^{-1}$ resulted in
$\gamma(BCES)=0.58\pm0.19$ and
$\gamma(fitexy)=0.74\pm0.04$. Reverberation mapping of multiple line
species in the same object and extending the luminosity range of
reverberation mapped sources would clearly be of great benefit in
eliminating some of uncertainties with this analysis.
  
\subsection{Narrow emission lines and black hole masses}

A number of studies have found that the stellar velocity dispersion of
the host galaxy bulge is related to the bulge luminosity which is, in
turn, related to the mass of the black hole (e.g. Magorrian et
al. 1998; Ferrarese et al. 2001; Tremaine et al.
2002)\nocite{Tremaine02}\nocite{Ferrarese01}\nocite{Magorrian98}.
Recently, Shields et al. (2002) and Boroson
(2003)\nocite{Shields02}\nocite{Boroson03} investigated the
possibility of using the velocity width of \oiii\ as a substitute for
the stellar velocity dispersion when determining the black hole
mass. They both found that the velocity width of \oiii\ is correlated
with black hole mass, assuming that the black hole mass $\propto$
$L^{\gamma}v^{2}$ where $L$ is the QSO luminosity at 5100\AA, $v$ is the
velocity width of the \hb\ emission line and $\gamma=0.5$ and 0.7
for Shields et al. (2002) and Boroson (2003) respectively. We find
that the velocity width of \oiii\ is correlated with the luminosity of
the QSO (Table 1) with $v\propto L^{0.1\sim 0.15}$ (Table 3; FWHM
values as the IPV values are probably contaminated by \feii\ emission
either side of the \oiii\ lines), similar to that found for \hb. This
raises the intriguing possibility first, that the NLR in QSOs is
sufficiently close to the black hole for its motion to be dominated by
the gravitational potential of the black hole, and secondly, that the
narrow emission lines may be used to derive the black hole
mass. However, none of the other narrow emission lines in our study
(\oii, \nev\ and \neiii) display the same strength of correlation
between line width and QSO luminosity as \oiii\ and given the
anti-correlation between narrow line strength and continuum luminosity
(Croom et al. 2002) the use of narrow lines in estimating the black
hole mass may be limited to low luminosity and hence low mass objects.

\subsection{Evolution of the black hole mass-to-luminosity ratio}
 
Figure \ref{vel_bm_noscal} shows the mass--luminosity relation for the
individual broad lines separated into redshift bins and appears to
demonstrate a lack of evolution in the relation between black hole
mass and luminosity with redshift.  However, the \civ\ line does show
a significant anti-correlation between velocity and redshift in our
partial correlation analysis (Table 2) which may be evidence of a weak
redshift evolution of line width. No other lines show significant
correlations with redshift.
  
In order to make a quantitative estimate of the amount of redshift
evolution we fit a simple power law model to the $L-z$ data points
for each emission line.  We first assume that black hole mass is
related to luminosity and redshift by independent power laws such that
\begin{equation}
\mbh(z)\propto L(z)^\alpha(1+z)^\beta,
\end{equation}
and that secondly, the radius-luminosity relation is also described by
a power law:
\begin{equation}
R\propto L(z)^\gamma.
\label{eq:rlevol}
\end{equation}

We assume that the radius-luminosity relation (Equation
\ref{eq:rlevol}) does not to
evolve with redshift as it should be primarily driven by the physics
of photoionization, which we would not expect to change with redshift
(although this may not be the case if QSO metallicity was found to
evolve).  Assuming Keplerian motion, we have
\begin{equation}
\log(v)=\frac{\alpha-\gamma}{2}\log(L)+\frac{\beta}{2}\log(1+z)+Const.
\end{equation}
 
\begin{figure}
\centering \centerline{\psfig{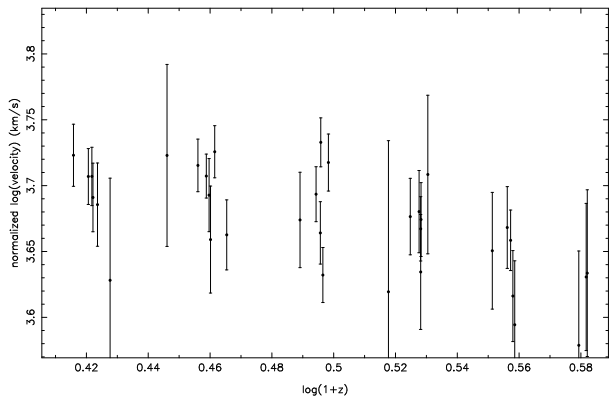}}
\caption{The evolution of \civ\ velocity width.  We plot the
re-normalized (to a fixed absolute magntude of $\mb=-25.0$) velocity
vs. $(1+z)$ in log-log space. The error bars have been re-scaled such
that $\chi^2$ is equal to the number of degrees of freedom.  A decline
in velocity can be seen at the highest redshifts.}
\label{fig_civ_evol}
\end{figure}

This relation is then fitted to the observed velocity-luminosity data
for each emission line individually (as we do not wish to assume that
the radius-luminosity relation is the same for each emission line).
To obtain a more realistic estimate of the errors on the fitted
parameters given a large intrinsic scatter in the data, we re-scale
the errors on individual points such that $\chi^2$ is equal to the
number of degrees of freedom.  The results of this model fitting are
shown in Table \ref{table_evol}.  The only line which shows clear
evidence ($>3\sigma$) for evolution is \civ.  However, \siiv\ and
\ciii\ also show weak evolution in the same sense.  All of these lines
have best fit values of $\beta$ which are negative, implying that for
a given luminosity, the measured velocities are lower at higher
redshift.  By contrast, the value of $\beta$ for the \mgii\ line is
entirely consistent with no evolution at all and the Balmer lines
(\hb\ and \hg) all show positive values of $\beta$, although these are
not significant. We do not attempt to model the redshift evolution of
the \lya\ line, as the contribution of \lya\ forest absorption and
\nv\ emission makes any interpretation highly uncertain.  To show more
clearly the evolution in the \civ\ velocity width, we plot $\log(v)$
vs. $\log(1+z)$ for this line, first renormalizing the velocity points
to a constant absolute magnitude (based on the above fits) of $\mb=-25.0$
(Fig. \ref{fig_civ_evol}).  The error bars in Fig. \ref{fig_civ_evol}
have been rescaled such that $\chi^2$ is equal to the number of
degrees of freedom.  A weak, but significant, decline in velocity can
be seen, particularly at $\log(1+z)>0.5$.

\begin{figure}
\centering \centerline{\psfig{file=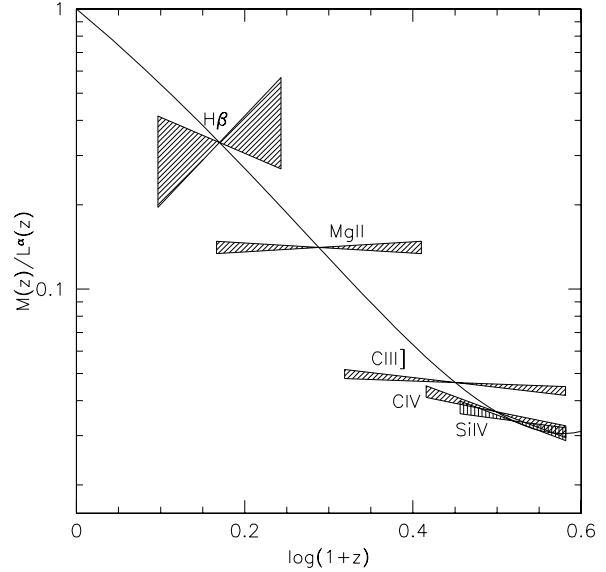,width=8cm}}
\caption{The best fit evolutionary power law relationships.  The
evolution predicted by the luminosity evolution of Boyle et al. (solid
line) is compared to the slopes found in our fits (shaded
regions). The horizontal width of the individual shaded regions
represents the redshift range covered by our data for each emission
line and the height of the regions represents the $1\sigma$ error on
$\beta$ (Table 5). The vertical positioning of the shaded regions is
such that they intersect the model at their mean redshift (see text
for a more detailed discussion). } 
\label{fig_evol_slopes}
\end{figure}

The evolution seen in the higher redshift lines is in the sense of
higher redshift QSOs having smaller velocities for the same
luminosities, which might be expected if QSO luminosity evolution is
driven by a decline in fueling rate at lower redshift.  However, the
amount of evolution is small compared to the strong luminosity
evolution shown by QSOs (typically $\sim(1+z)^3$).  At $z\sim2-3$ the
QSO evolution slows down and appears to turn over.  Assuming pure
luminosity evolution (PLE), Croom et al. (2003) finds that for the
full 2QZ catalogue the evolution of $L^*$, the characteristic
luminosity of the QSO population, is well described by
$L^*(z)=L^*(0)10^{k_1z+k_2z^2}$ with $k_1=1.17$ and $k_2=-0.21$ at
redshifts less that $z\sim2.3$. This model implies that $L^*$ reaches
an extrapolated maximum at $z\sim2.8$.  If black hole masses do not
evolve, and the luminosity evolution is purely due to other influences
(e.g. fueling rate), then at a given luminosity $\mbh(z)=\mbh(0)$.  Thus,
using the PLE fit from Boyle et al (2003) and assuming that $M\propto
L^{\alpha}$ we find
\begin{equation}
\frac{\mbh(z)}{L^{\alpha}(z)}=\frac{\mbh(0)}{L^\alpha(0)}10^{-\alpha(k_1z+k_2z^2)}.
\end{equation}
In Figure \ref{fig_evol_slopes} we compare this model to the
evolutionary slopes found in our fits. Our PLE model (solid line) is
steep at low redshift, but flattens off at $z\sim2-3$.  Here we assume
that $\alpha=0.93$, as found in Section 5.1 for all broad emission
lines, but the shape of the model curve changes little if
$\alpha=1.0$, as found for the \hb\ and \civ\ lines alone (Section
5.1).  We plot, as shaded regions, the range of allowable
($\pm1\sigma$) slopes for each of five lines (\hb, \mgii, \ciii, \civ\
and \siiv), over the redshift ranges that they are measured.  Each is
normalized to the $M(z)/L^\alpha(z)$ ratio at the mean redshift of the
line.  At low redshift, the \hb\ line has a positive slope which is
consistent with $\beta=0$, but not with the strong evolution of the
model (solid line).  At higher redshift the \mgii\ and \ciii\ lines
are close to $\beta=0$ and although this is inconsistent with our
model, we note that both these lines are contaminated by blended
emission which could be distorting their behaviour. Only the highest
redshift lines, \civ\ and \siiv\ show evolution which is comparible to
that shown by the model, partly as they have higher values of $\beta$
and partly as the model starts to turn over at high redshift.

\begin{table*} 
\caption{The best fit evolutionary power law slopes.}
\label{table_evol}
\begin{center}
\begin{tabular} {lrr}
\hline Line & $\beta$ & $\Delta\beta$ \\ 
\hline 
\siiv\   & $-$0.76 & 0.36 \\ 
\civ\    & $-$0.86 & 0.25 \\ 
\ciii\   & $-$0.23 & 0.12 \\ 
\mgii\   &    0.00 & 0.18 \\ 
\hg\     &    0.84 & 1.85 \\ 
\hb\     &    0.94 & 2.23 \\ 
\hline
\end{tabular}
\end{center}
\end{table*}

\begin{figure*}
\centering \centerline{\psfig{file=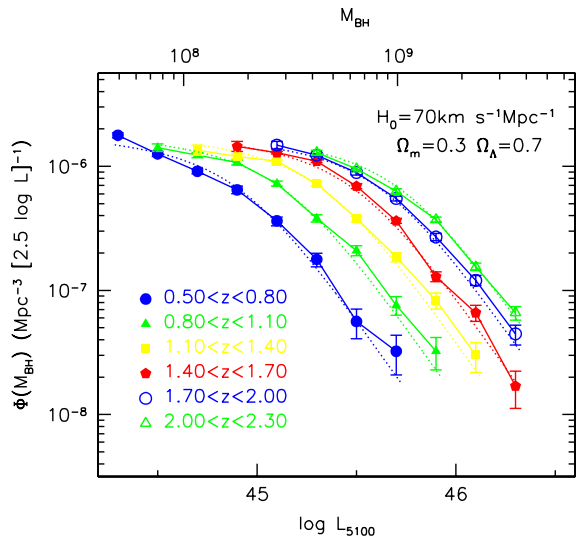,width=8cm}}
\caption{The Black Hole Mass Function for QSOs and its evolution with redshift.}
\label{bhmf}
\end{figure*}

It appears that only at $z>2$ does the measured evolution in the
luminosity-velocity relation compare to the above model.  Over the
majority of the redshift range covered in this investigation evolution
in the luminosity-velocity relation is insignificant.  This conclusion
has significant implications for QSO evolution, implying little
evolution in the mass-luminosity relation for super-massive black
holes over a wide range in redshift.  Assuming no evolution in the
mass-luminosity relation we can transform the QSO luminosity function (LF)
and its evolution with redshift into a mass function (MF).  In
Fig. \ref{bhmf} we show the evolution of the QSO mass function based
on the LF derived from the 2QZ+6QZ samples (Croom et al.\ 2003).  We
immediately note that the luminosity (mass) evolution of the QSO LF
cannot be explained by decline in the fueling rate -- either in
continuous model (long-lived QSOs) or in a duty cycle (short-lived
QSOs), since this would imply that a QSO at a given relative position
on the LF (MF) evolving from high black hole mass at high redshift to
low black hole mass at low redshift.
 
Rather, the evolution of the black hole mass function suggests that
QSOs can only be active once, possibly associated with formation of
the black  hole itself during spheroid formation. Under this
interpretation, the evolution of the LF is driven by the efficiency of
black hole formation which drops rapidly towards low redshift.  This
picture is qualitatively consistent with the recent theoretical work
of Kauffmann \& Haehnelt (2000)\nocite{Kauffmann00}, although it
places tight constraints on the fraction of QSOs which exhibit a duty
cycle of activity in such models.

It also implies that the Gebhardt et al.\ (2002) relation between
black hole mass and galaxy velocity dispersion will evolve toward
higher ratios at higher redshifts.  The black hole mass will be
'frozen' when black hole formation and subsequent QSO activity ceases
while the mass/velocity dispersion of the host galaxies will increase
due to hierarchical merging to give the relationship seen at the
present day.

At $z\sim2$ and greater, there does seem to be some evolution of the
velocity-luminosity relation, as shown by the \civ\
(Fig. \ref{fig_civ_evol}) and \siiv\ lines.  It appears that there is
a general decline in velocity for a given luminosity at
$\log(1+z)>0.5$ ($z>2.2$).  The measured rate of evolution is no
different if we remove the highest redshift bin ($2.75<z<3.0$) and
re-fit the data.  The fact that the strongest evolution in the
velocity-luminosity relationship is found in the redshift regime
corresponding to the peak of QSO activity suggests that as $L^*$
declines at high redshift ($z>2.5$), the black hole mass-luminosity
relation may be changing more rapidly than at lower redshift.

\section{Conclusions}

We have used composite spectra generated from more than 22000 QSOs
observed in the course of the 2dF and 6dF QSO redshift surveys to
investigate the relationship between the profiles of emission lines
and luminosity.  We find that the velocity width of the broad emission
lines \hb, \hg, \mgii, \ciii\ and \civ\ show a positive correlation
with the continuum luminosity (as measured in the $\bj$ filter) but the
slope of this dependence is much flatter for the UV emission lines
(\mgii, \ciii\ and \civ\ ) than for the hydrogen lines. This
correlation is statistically significant with a confidence level 
$>99$ per cent and is independent of the method used to measure the
emission line widths. From partial Spearman correlation analysis we
find only weak evidence for an anti-correlation between redshift and
velocity. Of the narrow emission lines, \oii, \neiii, \nev\ and \oiii,
only the \oiii\ $\lambda$5007 line exhibits a correlation between line
width and luminosity. 

Assuming that the gas in the BLR is in near-Keplerian orbits and that
the radius of the BLR, $r \propto L^{0.68}$ (K00;Netzer 2003), where
$L$ is the continuum monochromatic luminosity at 5100\AA, we have used
our measurements of the emission line velocity widths to derive
estimates of the average black hole mass in the composite spectra. We
find that for a given luminosity the black hole masses obtained from
the different the emission lines agree to within 0.6dex. However the
scatter in the mass-luminosity relationships for the individual lines
is much smaller ($<0.3$ dex) and given that the different emission
lines may be emitted from different radii, producing an offset between
the masses estimated from different emission lines, we use the \hb\
$\mbh-L$ relationship to calibrate the other emission lines. We find
an overall black hole mass - luminosity relationship of the form
$\mbh\propto L^{0.93\pm0.05}$. It is possible that the FWHM 
measurements of several of the emission lines are overestimated as
they are blended with weaker lines from different elements which are
difficult to remove reliably. We have therefore also derived the
mass--luminosity relationship for the \hb\ and \civ\ emission lines
only, as these are the lines least effected by contaminating emission,
and find $\mbh\propto L^{1.0\pm0.1}$.  Our$\mbh-L$ relationships depend
heavily on the slope of the $\log(r)-\log(L)$ relationship which is
currently known to an accuracy of only $\pm0.15$. Using the midpoint
of the derived $\mbh-L$ powerlaws and taking into account the error in
the slope of the $\log(r)-\log(L)$ relationship, we conclude that
$\mbh\propto L^{0.97\pm0.16}$.

We find only weak evidence of evolution in the velocity-luminosity
correlation, which places strong constraints on the evolution of black
hole masses for a fixed QSO luminosity; $M\propto (1+z)^{\beta}$,
$\beta<1$. This is much smaller than that required for the luminosity
(mass) function of QSOs to be due to a single population of long-lived
sources.

\section*{Acknowledgments}

The 2dF QSO Redshift Survey was based on observations made with the
Anglo-Australian Telescope and the UK Schmidt Telescope.  We warmly
thank all the present and former staff of the Anglo-Australian
Observatory for their work in building and operating the 2dF facility.

\end{document}